\documentclass[10pt]{article}
\usepackage[]{aas_macros}
\usepackage[dvips]{graphicx}
\usepackage{icrctc07}
\title{Studies of Direct Cherenkov Emission with VERITAS}
\shorttitle{Studies of Direct Cherenkov Emission with VERITAS}

\authors{S. A. Wissel$^{1,2}$, for the VERITAS Collaboration$^{3}$ }
\shortauthors{S.A. Wissel et al.}
\afiliations{$^1$Kavli Institute for Cosmological Physics, Chicago, IL 60637, U.S.A.\\ $^2$Enrico Fermi Institute and Department of Physics, University of
  Chicago, Chicago IL 60637, U.S.A.\\
  $^3$For full author list see G. Maier, ``VERITAS: Status and Latest Results", these proceedings }
\email{wissels@uchicago.edu}

\abstract{
Ground-based composition measurements of high-energy cosmic rays can be
significantly improved by using the direct Cherenkov method. This technique
 targets the Cherenkov light produced by the primary particle prior to its
production of an extensive air shower. With the appropriate time and angular
resolution, the direct Cherenkov photons can be separated from those produced
in the extensive air shower. By utilizing the $0.15^{\circ}$ angular and 2 nanosecond
timing resolution of the very high energy gamma-ray telescope system, VERITAS,
the charge and energy of cosmic rays at TeV energies can be identified on an
event-by-event basis. Results from a preliminary search for direct Cherenkov events are discussed.}

\begin{document}
\maketitle
\section{ Background }
 One of the primary sources of high-energy cosmic rays up to the knee region (of energies $\sim10^{15}$eV) seems to be
galactic supernovae remnants, but there remain open questions as to how the cosmic rays are accelerated to such high energies.
Particles confined to the region close to a relativistic shock can undergo diffusive shock acceleration \cite{1978MNRAS.182..147B, 1977DoSSR.234R1306K} in which the maximum energy produced by supernova remnants is argued to be $Z\times10^{14}eV$\cite{1983A&A...125..249L}.

 Key to understanding galactic cosmic-ray acceleration and the transition to extragalactic sources is the break in
the spectrum at knee energies.  The elemental composition as a function of energy is particularly interesting,
because of the potential dependence of the cosmic-ray spectral break on cosmic-ray mass and abundance.  A recent
flight of the TRACER balloon experiment has measured the composition from neon to iron up to TeV energies \cite{2007astro.ph..3707B},
and the experiment, along with the recent H.E.S.S. iron measurements using the method described here
see no spectral break \cite{2007astro.ph..1766H}.
\section{Method}
 Before interacting strongly and producing the
well-known extensive air shower (EAS), a high-energy cosmic ray will
give rise to Cherenkov radiation in the surrounding atmosphere.
The flux of this radiation, the Direct Cherenkov (DC) radiation, is dependant on
the primary's charge and the properties of the medium surrounding it.
A measurement of the total energy in the entire EAS, coupled
with the charge measurement from the DC light, allows for
event-by-event composition studies.

The DC light will arrive closer to to the shower core and later in time
than the extensive air shower light.  This is because the DC photons that
are not obscured by the extensive air shower start their descent
higher in the atmosphere where the air density is thinner, so
the Cherenkov angle of the DC photons is smaller than
those of the EAS light.  See \cite{2001APh....15..287K}
for a detailed exposition of the method.

The Very Energetic Telescope Array System (VERITAS) is a gamma-ray telescope array designed for $100$GeV$-30$TeV gamma-ray studies. Since
March 2007, it has been running in operating mode
with 4 $12m$ Davies-Cotton telescopes at Mount Hopkins in Amado, Arizona. Two telescopes are 35m apart,
while the other two are placed 85m and 81m from the first two in such a way that the telescopes
form an asymmetric V-shape. Each telescope's camera consists of 499 circular single-anode photomultipliers,
covering a $3.5^{\circ}$ field-of-view. Custom 500 Msps FADCs digitize the phototube signals into events of 24 samples
when the event passes the pattern and array trigger requirements. The
angular pixel size of 0.15$^{\circ}$ and 2ns digitization are adequate for Direct Cherenkov studies.
See the discussion by G. Maier in these proceedings for a more details on the detector\cite{icrc2007_Maier}.
\subsection{VERITAS Timing Studies}
 A useful aspect of the VERITAS detector is its timing capabilities.  Because the separation in time between the extensive air shower
and the DC pulse is on the 2ns scale, both signals will arrive within one 48ns event. Since the Cherenkov emission from the primary that is created high in the atmosphere accumulates into only 1-3 pixels early in the shower, a measurement of the arrival times of the photons provides an extra degree of freedom that will prove fruitful.

\begin{figure}[ht]
\begin{center}
\noindent
\includegraphics [width=0.5\textwidth]{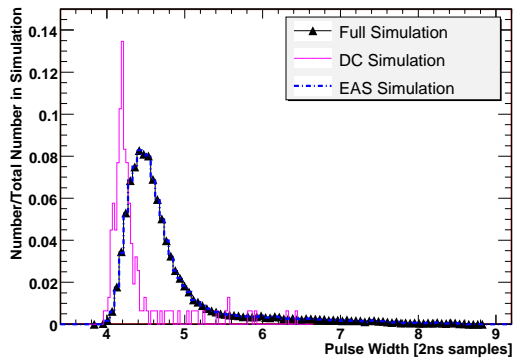}
\caption{Simulations of phototube signals from 748 10-500TeV $^{56}$Fe primaries.  The samples are divided according to the ratio of photons emitted above the first interaction height to those emitted below the first interaction height.  So that only the pixels near the beginning of the air shower are considered, a minimum arrival time cut of 14ns has been applied.  The arrival time, as measured with respect to the time of the array trigger,  is defined as the time at which the PMT trace reaches its full-width half-maximum. These simulations also require that each pixel have an integrated charge sum of 1000 digital counts and that the telescope is between 60m and 90m from the shower core. \label{icrc0732_fig01}}
\end{center}
\end{figure}
 For instance, sufficiently high-$Z$ cosmic rays produce a large, focused beam of Cherenkov radiation, so the full-width half-maximum (width) of the pulse in an individual photomultiplier tube is generally narrower than the total width of the air shower.  A full detector simulation of high-energy cosmic rays has been performed to evaluate the timing characteristics of such showers in the VERITAS array.  Here, CORSIKA 6.5021 simulations \cite{CORSIKA} of 748 10-500 TeV iron showers were divided into two populations in which the phototubes in the Direct Cherenkov sample have at least 50\% of their photons emitted above the first interaction height, while the air shower sample contains the rest of the pixels in the image. Pulse width distributions show that while the focused Cherenkov beam from the primary is smoothed by the intrinsic spread of the photomultiplier tubes, the spread in pulse widths is smaller than that from the air shower photons (Figure \ref{icrc0732_fig01}).
\section{ Preliminary Search }
 A preliminary search for DC emission in VERITAS is currently underway for the Fall 2006-Spring 2007 observing season.
The effective area imposed by annulus cuts allows only $\sim1$ observable
100 TeV $^{56}$Fe event per night, so optimizing the cuts for high-energy cosmic rays is crucial to studies such as these.
In addition to quality cuts on Hillas parameters, there are three strict requirements being used for event selection: an accumulation cut, an annulus cut, and a size cut. If the two-dimensional shower image is divided into quarters along the major-axis of an ellipse by lines parallel to the minor axis, then the ``front" of the image includes the pixels which lie within the first quarter closest to the shower core location and the ``back" of the image includes the pixels in the rest of the image.  The accumulation cut mandates that the front be twice as bright as the back.  The annulus cut requires that the position of the core with respect to the telescope position remain within a 60-90m radius, to maximize the DC photon yield.  In the absence of an EAS, the Cerenkov emission from the primary would be greatest near the maximum geometrically allowed radius, because there is a photon pileup due to the increasing emission angle as emission height decreases. However, since the cosmic ray will interact hadronically, the best observable impact parameter is a compromise between the region where there is the highest DC photon yield and the regions where the EAS dominates.  The size cut dictates that the sum of the total integrated charge in an image be greater than 1000 digital counts.

\begin{figure*}
\begin{center}
\includegraphics [width=1.1\textwidth]{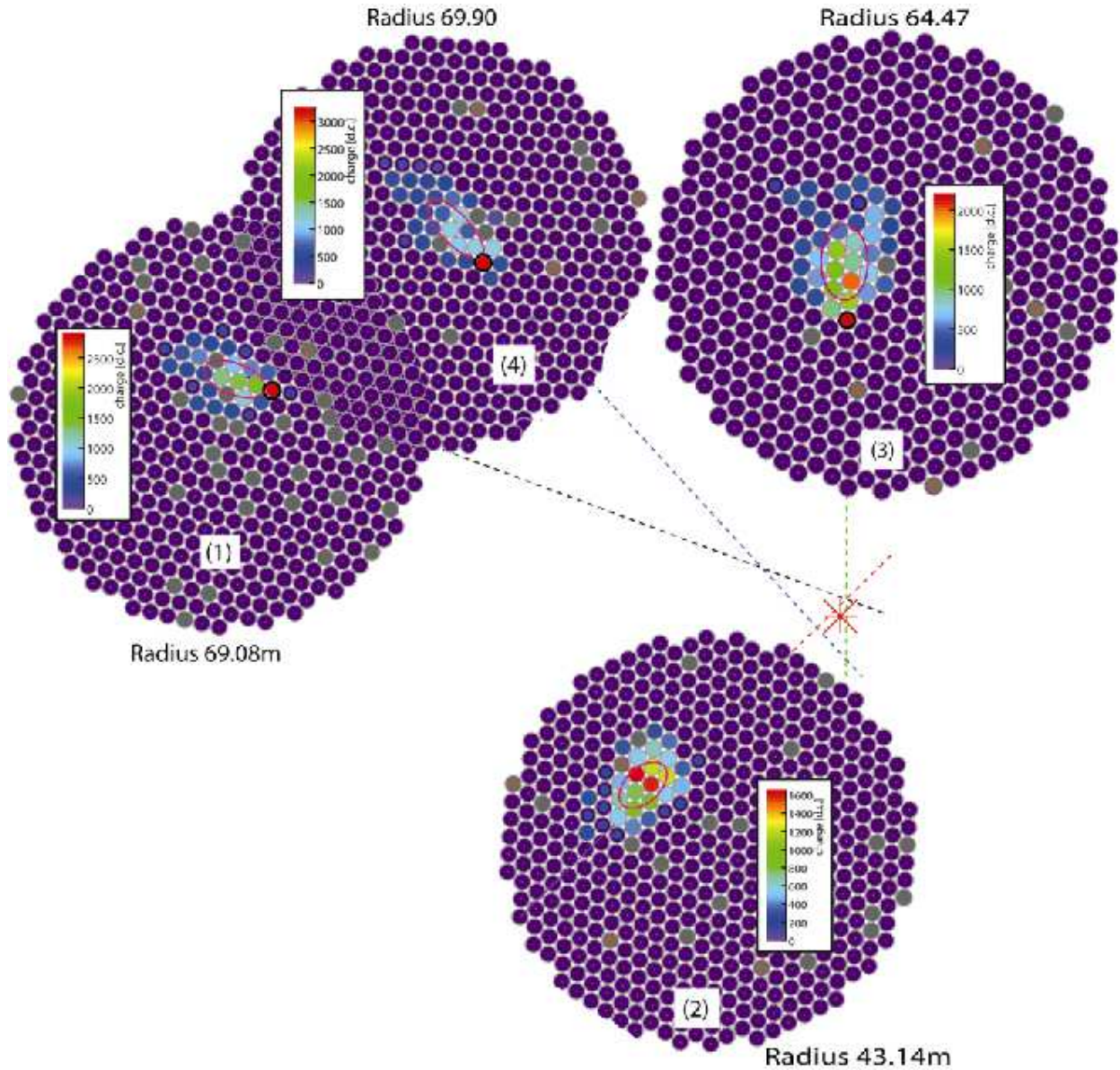}
\caption{A candidate event displayed as integrated charge sums on the camera planes. When a telescope is too close to the shower core($<50$m as in telescope 2), the density of primary Cherenkov photons is too low to produce a significant DC excess, but at radii between 60 and 70 meters (as in telescopes 1,3, and 4), the maximum number of DC photons seen by each of the three telescopes is ~500 photoelectrons.
   }\label{icrc0732_fig02}
\end{center}
\end{figure*}

 A 4-telescope candidate event, shown in Figure \ref{icrc0732_fig02}, demonstrates a typical Direct Cherenkov event in VERITAS. The DC peak of $\sim500$ photoelectrons is seen in telescopes 1, 3, and 4 as the bright pixel closest to the shower core position, while in the second telescope the brightest pixel is where one would expect the EAS center-of-gravity to be. The unique telescope configuration of VERITAS may assist in identifying the DC pulse through 3 out of 4 telescope coincidences, because the DC pulse could arrive at the same time and shower angle in 2 cameras that are close to each other as well as in one camera that is further away, but not in one that is at a radius in which the DC photon yield is low.  Studies into the effects of the asymmetric baseline, as well the arrival of cosmic-ray events in both time and angle, are currently being performed.
\subsection{ Discussion }
 Along with other imaging atmospheric Cherenkov telescopes, VERITAS has the capabilities to use the Direct Cherenkov method to identify the mass and energy of cosmic rays at energies that necessitate large effective areas.
 Because the flux at high energies is low, utilizing the timing information from VERITAS will help accurately identify Direct Cherenkov events as is being done in the preliminary studies presented here.  Future dedicated Direct Cherenkov detectors, such as the successor to the Track Imaging Cerenkov Experiment (TrICE) \cite{icrc2007_Hays} will benefit from improved angular and timing resolution.
\section*{Acknowledgments}
This research is supported by grants from the U.S. Department of Energy,
the U.S. National Science Foundation,
and the Smithsonian Institution, by NSERC in Canada, by PPARC in the UK and
by Science Foundation Ireland.
\bibliography{icrc0732}
\bibliographystyle{unsrt}
\end{document}